# The Use of Color Sensors for Spectrographic Calibration

Neil B. Thomas
*Department of Astronautical Engineering, United States Air Force Academy, CO 80840, USA*

**The wavelength calibration of spectrographs is an essential but challenging task in many disciplines. Calibration is traditionally accomplished by imaging the spectrum of a light source containing features that are known to appear at certain wavelengths and mapping them to their location on the sensor. This is typically required in conjunction with each scientific observation to account for mechanical and optical variations of the instrument over time, which may span years for certain projects. The method presented here investigates the usage of color itself instead of spectral features to calibrate a spectrograph. The primary advantage of such a calibration is that any broad-spectrum light source such as the sky or an incandescent bulb is suitable. This method allows for calibration using the full optical pathway of the instrument instead of incorporating separate calibration equipment that may introduce errors. This paper focuses on the potential for color calibration in the field of radial velocity astronomy, in which instruments must be finely calibrated for long periods of time to detect tiny Doppler wavelength shifts. This method is not restricted to radial velocity, however, and may find application in any field requiring calibrated spectrometers such as sea water analysis, cellular biology, chemistry, atmospheric studies, and so on. This paper demonstrates that color sensors have the potential to provide calibration with greatly reduced complexity.**

## 1. Introduction

The spectral analysis of light is fundamental to many fields of research. Environmental research, medical diagnostics, and especially astronomy rely heavily on the information contained in light at its various wavelengths. Such research often requires that absorption or emission features be identified and that their changes in shape or wavelengths are tracked over time. Doing so demands that each location on the recorded spectral image has a known correspondence to wavelength. Creating such a wavelength solution is often accomplished by observing a light source that is rich in features and then matching these known wavelengths to their apparent position on the sensor to determine wavelength as a function of location, λ(x). Performing this calibration with each scientific observation allows for precise wavelength measurements even when the instrument significantly changes due to environmental, mechanical, or optical disturbances.

This paper focuses on the field of astronomy and the use of spectroscopy to detect the time-varying Doppler shifts in stellar light caused by line-of-sight radial velocity (RV) oscillations that are induced by the gravitational influence of orbital companions. This application is chosen because it is of particular interest to the author and is also a field in which precise calibration is essential. But the method presented here is not restricted to this application. RV is an extremely challenging application because the Doppler shifts induced by sub-stellar companions are very small. Specifically, the change in wavelength, Δλ due to RV is given by Equation 1, where c is the speed of light and $\lambda_{rest}$ is the rest wavelength of the observed feature. RV corresponds directly to Doppler shift in the case of an ideal measurement. Errors in wavelength measurement due to miscalibrations and noise, however, will directly lead to errors in the calculated RV

$$\Delta\lambda = \frac{RV\,\lambda_{rest}}{c} \tag{1}$$

The influence of a giant planet is typically only 100 $ms^{-1}$ or less, imparting a wavelength shift of order $10^{-3}$ Å in visible wavelengths. Spectrometers are not inherently stable to this level of precision and their calibration becomes the limiting factor in this field (Wright 2017).

The quest for improved precision has focused on calibration sources and intrinsic instrument stability (Wright 2017). Maintaining repeatable precision over long periods of time is the critical element in this field because changes in RV relative to a reference epoch provide the information necessary to characterize the mass and orbit of a companion. Absolute accuracy may be required for certain research efforts, however, such as the study of stellar kinematics within our Galaxy (e.g. Grieves et al. 2017).

The most popular calibration sources currently employ an Iodine gas absorption cell (e.g. Marcy & Butler 1992) or a Thorium-Argon (ThAr) lamp (e.g. Mayor et al. 2003). The calibration exposures can be captured in sequence with the primary image or collected simultaneously by passing the starlight through an Iodine cell or dedicating a second optical path for ThAr emissions. Processing is simpler using separate exposures but differences between the stellar and calibration paths often introduce errors (e.g. Thomas et al. 2016). Combined exposures ensure that calibration information is simultaneous with the stellar exposure, but the separation of calibration and stellar features is complicated. Gas cell absorption methods have more or less stagnated at the 1 – 10 $ms^{-1}$ level of precision (e.g. Butler et al. 1996). Newer instruments now push below 1 $ms^{-1}$ precisions, primarily using ThAr (e.g. Queloz et al. 2001). Even so, emission sources have fundamental limits. Since only the spectral features provide information (Bouchy et al. 2001), the source must provide many well-distributed features that cover the instrument's wavelength range without being overly blended. Improving the inherent stability of the instrument eases the calibration burden, but controlling the environment to the necessary levels becomes a primary design consideration (e.g. Mayor et al. 2003). Fabry-Pérot etalons have performed comparably to or slightly better than ThAr, their primary advantage being a larger wavelength coverage and regularly spaced calibration features (e.g. Wildi et al. 2010). Future instruments such as EXPRESSO (Pepe et al. 2010) intend to take the next step in precision through the use of laser frequency combs. While offering the potential of $cms^{-1}$-level precision, laser combs are even less available to the general community. Alternatively, multi-object surveys such as LAMOST (Wang & Luo 2012) and MARVELS (Ge et al. 2009) produce highly useful science with accuracies in the 100 $ms^{-1}$ to 1 $kms^{-1}$ range. The analysis and simulation presented in this work demonstrate an approach to using commercially available color sensors to achieve scientifically viable precisions without the need for dedicated calibration equipment or specialized light sources, except during initial instrument characterization. The ultimate limit of this approach will require experimentation. But the potential is shown for much simpler, lighter, and less expensive spectrographic calibration.

Astronomers almost universally rely on monochrome sensors for a variety of good reasons. Color filters, whether they cover the entire optical pathway or are part of each pixel, reduce the overall number of photons recorded by the sensor. Additionally, the simulation of color requires the use of multiple pixels and reduces the effective spatial resolution. But we know that the color of light is synonymous with its wavelength and that the color of the light itself betrays its wavelength. The method detailed in this paper uses the mix of colors detected at each location on the sensor to determine the wavelength solution.

In Section 2 the methodology for using data from a color sensor is developed to determine wavelength, including dominant sources of error. A prototypical application is presented in Section 3 by simulating the characteristics of an off the shelf (OTS) color sensor and broadband calibration light source. Section 4 discusses areas for improvement, limitations, and suitable applications. Section 5 is the conclusion.

## 2. Methodology

Modern image sensors simulate color photography by covering pixels with arrays of red, green, and blue filters in a pattern called a Bayer matrix. The arrangement of Bayer matrix filters on the pixels of a sensor is illustrated in Figure 1. The information from four pixels is used to create the true color of that region. The color at a location in the image is synthesized based on the relative intensities of nearly collocated pixels.

| | | | |
|---|---|---|---|
| Blue | Green | Blue | Green |
| Green | Red | Green | Red |
| Blue | Green | Blue | Green |
| Green | Red | Green | Red |

*FIG. 1. The Bayer matrix arrangement of color filters on the pixels of a sensor. The observed brightness in red, green, and blue is used to determine the color in a region. Interpolation algorithms are used to create an entire color image appearing to have the spatial resolution of the underlying sensor.*

The quantum efficiency (QE) for a filtered pixel is the ratio of photons detected by the pixel to the number entering the filter and is a function of wavelength. This is slightly different than the meaning of QE for monochromatic sensors, in which QE is strictly a function of a pixel's ability to convert photons into electrons ($e^-$). The QE of a pixel in a color sensor includes the influence of its filter. The QE will, at best, approach that of the unfiltered version within a limited wavelength range and is intentionally much lower elsewhere. Manufacturers generally provide QE(λ) for each color in a form such as Figure 2, with all pixels of a given color on the sensor assumed to have the same response. Commercial sensors commonly use red, green, and blue filters to capture the information necessary to recreate the human perception of color in a final image, although this selection of colors is somewhat arbitrary from a physics standpoint. While the QE performance of the three filter types is often presented in one figure for simplicity, they are independent from one another. All references to QE(λ) throughout this paper imply a pixel's efficiency in combination with its color filter.

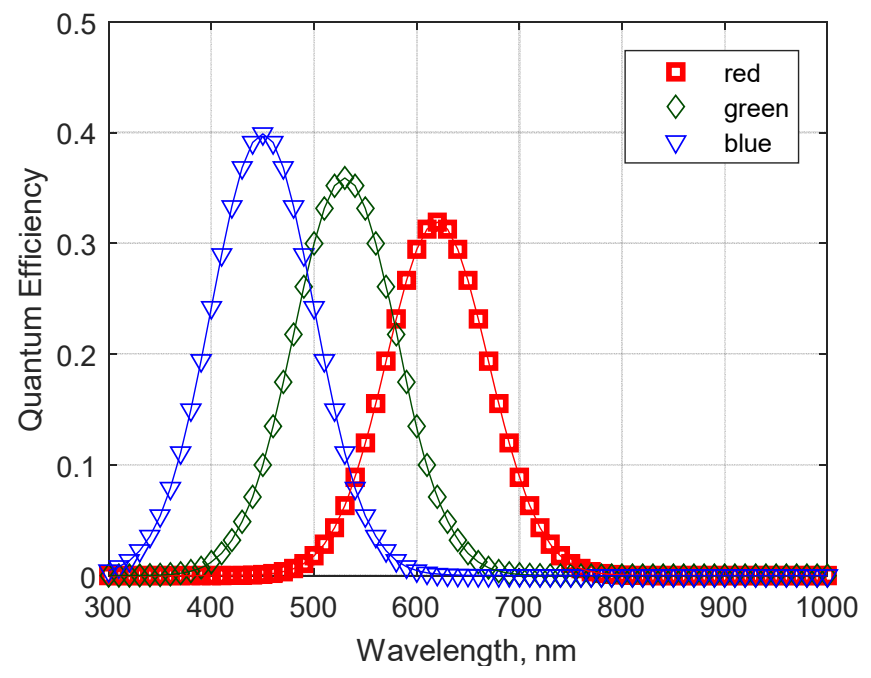

*FIG. 2. A simplified example of quantum efficiencies as a function of wavelength for the red, green, and blue filtered pixels of a color sensor.*

Since QE is a smooth function of λ, obtaining a measurement of QE allows the determination of λ within any monotonically increasing or decreasing region of QE(λ). Unfortunately, a single filter cannot provide QE unless the absolute flux is known, which is rarely the case. But when a group of pixels having N filter types (colors) are exposed to light of the same wavelength and intensity then the color fraction, F, of photons collected as electrons for a given color is given by Equation 2. The subscripts denote the filter type, either C for the color of interest or n for any color contributing to the sum of the signal being collected at the specified wavelength.

$$F_C(\lambda) = \frac{QE_C(\lambda)}{\sum_{n=1}^{N} QE_n(\lambda)} \tag{2}$$

This fraction can be calculated as a function of wavelength for each filter type using QE(λ) data from the manufacturer or as found separately by experimentation. Each color will have its own function which is dependent only on the sensor's color characteristics and not on the spectrograph or the target light source. Changes in the spectrograph may lead to variations in the photon throughput or the alignment of the projected spectrum, but the fractional composition of colors for a specific wavelength will remain constant, eventually allowing for the determination of a wavelength solution. Figure 3 shows $F_C(\lambda)$ for red, green, and blue colors after applying Equation 2 to the respective QE(λ) functions reported by the manufacturer of the KAI-11002[1] color sensor, which is selected for this discussion simply because it is representative of readily available sensors and has well-documented performance data. Unlike quantum efficiency, color fraction is a measure of the relative intensity of a color at the dispersion location. Color fractions are invariant to flux intensity and this allows for an observed value to be directly related to a specific wavelength.

---

[1] http://www.datasheet4u.com/datasheet-pdf/ONSemiconductor/KAI-11002/pdf.php?id=1077786

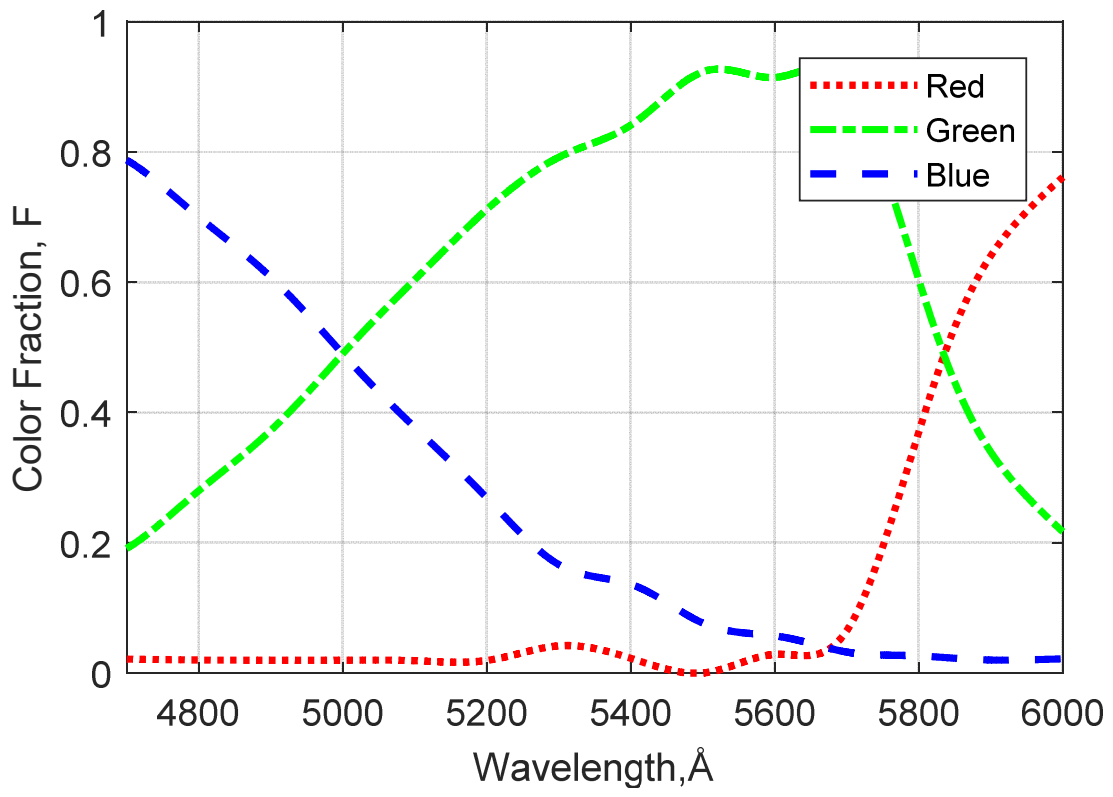

*FIG. 3. The color fraction for each color as a function of wavelength for the KAI-11002. The signal detected by a single color can be related to the light's wavelength. For instance, if the green signal accounts for 80% of the total signal then the wavelength is approximately either 5350 Å or 5750 Å, with the choice being obvious when restricted to a reasonable range.*

The entire procedure of matching the observed QE(x) to the sensor characteristics to determine wavelength hinges on a precise understanding of the sensor's QE(λ). The manufacturer generally provides this information, but far too crudely for these purposes. The color sensor's response will initially need to be determined using traditional calibration techniques. For instance, a medium to high resolution spectrometer may be used to disperse a well-studied light source such as ThAr. Each spectral feature having a known wavelength is then identified and the intensities reported by differently colored pixels are used to determine $F_C(\lambda)$. This can ideally be accomplished with the collection of a single spectrum that projects the necessary wavelength range onto the sensor assuming all pixels of the same color are identical. Any significant variation among pixels, due either to disparities in the pixels themselves or in the filter material, will be a source of calibration error. In that case, a more involved characterization of each pixel will be required, which is beyond the scope of this paper. The resolution of the calibration spectrometer ultimately drives the precision of the characterization. Although traditional calibration methods are initially required here, this does not detract from the novel advantages of the color calibration method. This sensor characterization is used to quantify the global response of the various colors of the sensor. Once completed, it is independent of future mechanical or optical changes. There is no requirement that the spectrometer used for characterization be related to the one used for scientific observations. Further, even gross mechanical changes, such as a macroscopic shift of the sensor, have no effect on the color response of individual pixels. This sensor characterization only needs to occur once and may be separate from the observational usage of the sensor if filter performance changes due to age and the environment are negligible. This type of filter stability is assumed in this paper but may need to be validated in practice.

Once the spectrograph is coupled to the color-characterized sensor for an observation, the Analog/Digital Converter (ADC) translates the number of electrons collected by each pixel into Analog/Digital Units (ADUs). The ADU reported by a pixel is given by Equation 3, where S is the signal measured in number of photons reaching the pixel's filter and g is the gain, which relates the number of electrons collected by the pixel to the reported ADU. The measured ADU is rounded due to the digital nature of the ADC.

$$ADU = round\left(\frac{(QE)\,S}{g}\right) \tag{3}$$

The color fraction for a given filter at location x along the dispersion direction, $F_c(x)$, is written in terms of ADU in Equation 4, noting that gain and signal divide out when gain is constant for all pixels and when the differently colored pixels at location x receive the same input flux.

$$F_C(x) = \frac{ADU_C(x)}{\sum_{n=1}^{N} ADU_n(x)} \tag{4}$$

The color fraction is based solely on the ADU counts reported directly by the sensor at the completion of the observation and requires no other information. This is similar to Equation 2 except that color information is obtained as a function of location not wavelength. Our strategy is to match observed $F_c(x)$ to the known sensor response $F_c(\lambda)$ to develop the instrument's wavelength solution, $\lambda(x)$.

The precision of the calibration will be limited by the precision of the flux measurements. The primary sources of noise in a pixel's measurement are photon shot noise, read noise, and dark noise. Shot noise is unavoidable and arises due to the discrete nature of photons arriving to each pixel. It appears as a normal distribution of uncertainty that is the square root of the signal created by photons, as measured by the number of electrons collected in a pixel. Read noise is introduced during the collection and counting of electrons at the end of an exposure. It generally introduces a constant amount of uncertainty to each pixel for a given sensor. Dark current adds to the electron count through the gradual collection of thermal electrons within each pixel during an exposure. Dark current accumulates predictably, but an associated dark noise is introduced in the same statistical manner as shot noise from the target signal. Dark current is modeled as a number of electrons per second and varies with sensor temperature, from which the uncertainty due to dark noise can be calculated. These independent sources of noise add in quadrature such that total noise, $\sigma_{total}$, for a reading is given by Equation 5, where TE is the number of electrons collected from the target signal, RN is the uncertainty in electron count introduced by read noise, and DC is the number of electrons introduced per unit of time by dark current.

$$\sigma_{total} = \sqrt{TE + RN^2 + (DC)time} \tag{5}$$

Dark noise may be neglected here because the calibration exposures can be kept quite short by making the source as bright as necessary. Read noise is generally on the order of 10 $e^-$ for low performance sensors, while the $e^-$ well depth typically[2] ranges from 10,000 to over 500,000 $e^-$. The $e^-$ well depth is the maximum number of electrons that can be stored in each pixel. Read noise is therefore an insignificant component of uncertainty compared to the signal except in very faint regions of an image. It is neglected here since it is assumed that the calibration signal will be matched to the much larger dynamic range of the sensor. It is, however, included in the later simulation for the sake of completeness. Since the signal captured from target photons is given by the product of the quantum efficiency of a filtered pixel and the number of available photons, the number of electrons collected in a pixel after an exposure is given by Equation 6.

$$e^- = \left(QE_C(x)S(x) \pm \sqrt{QE_C(x)S(x)}\right) \tag{6}$$

The ADU count reported by each pixel also includes rounding errors from the ADC's translation from $e^-$ to ADUs. The measured ADU for a colored filter as a function of the sensor's ADC characteristics and its shot noise is given by Equation 7.

$$ADU_C(x) = round\left[\frac{1}{g}\left(QE_C(x)S(x) \pm \sqrt{QE_C(x)S(x)}\right)\right] \tag{7}$$

Rounding error is negligible when compared to shot noise except for spectral regions of very low flux. Retaining the shot noise terms for each color but neglecting rounding, Equations 7 and 4 provide the predicted color fraction as a function of QE(x) values for each filter in Equation 8. The summation of noise terms in the denominator are combined in quadrature.

---

[2] http://www.clarkvision.com/articles/digital.sensor.performance.summary/index.html

$$F_C(x) = \frac{QE_C(x) \pm \sqrt{QE_C(x)/S(x)}}{\sum_{n=1}^{N} QE_n(x) \pm \sqrt{\sum_{n=1}^{N} QE_n(x)/S(x)}} \tag{8}$$

The awkward arrangement of errors in Equation 8 can be simplified using the error propagation property that the function, $f = A/_B$, having $\sigma_A$ error of the numerator and $\sigma_B$ error of the denominator, has an effective uncertainty given by Equation 9, where $\sigma_{AB}$ is the covariance between A and B.

$$\sigma_f \approx |f| \sqrt{\left(\frac{\sigma_A}{A}\right)^2 + \left(\frac{\sigma_B}{B}\right)^2 - 2\frac{\sigma_{AB}}{AB}} \tag{9}$$

Applying this to Equation 8 yields Equation 10, which provides a more intuitive representation of expected color fraction errors, $\sigma_F$, as a function of photon signal strength and filter characteristics.

$$F_C(x) \cong F_C(\lambda) \pm \frac{QE_C(x)}{\sqrt{S}\sum_{n=1}^{N} QE_n(x)} \sqrt{\frac{1}{QE_C(x)} + \frac{1}{\sum_{n=1}^{N} QE_n(x)}} = F_C(\lambda) \pm \sigma_F \tag{10}$$

This expression is approximate because it overestimates error by neglecting the covariance between the numerator and denominator in Equation 8 since $QE_C$ appears in both terms. This equation is the fundamental result of this work because it allows for ADU measurements at each dispersion location to be matched to a specific wavelength with a predictable level of uncertainty.

The color fractions are invariant to the flux of incoming light, except as they relate to the noise contributions. The technique is therefore relatively insensitive to the form or spectral characteristics of the calibration source. The illumination source must simply provide flux across the wavelength domain of interest. In fact, absorption features only serve to reduce the flux at certain wavelengths, decreasing the signal to noise ratio (SNR) and degrading the overall quality of the calibration. The continuum across this domain does not need to be especially flat, but a large variation in intensity should be avoided because it will systematically reduce the quality in lower flux regions and challenge the dynamic range capabilities of the sensor. Figure 4 demonstrates how a realistic observation can provide the information necessary to determine the color fractions. The top panel shows the incident light from a black body source being multiplied into the sensor's QE(λ) to generate the red, green, and blue signals recorded by the sensor, to include shot and read noise. The bottom panel is the same data represented in color fraction form. This is directly comparable to Figure 3 except that the observation is based on pixel location instead of wavelength, allowing for the development of the sensor's wavelength solution by finding the wavelength at which F(λ) is equal to F(x).

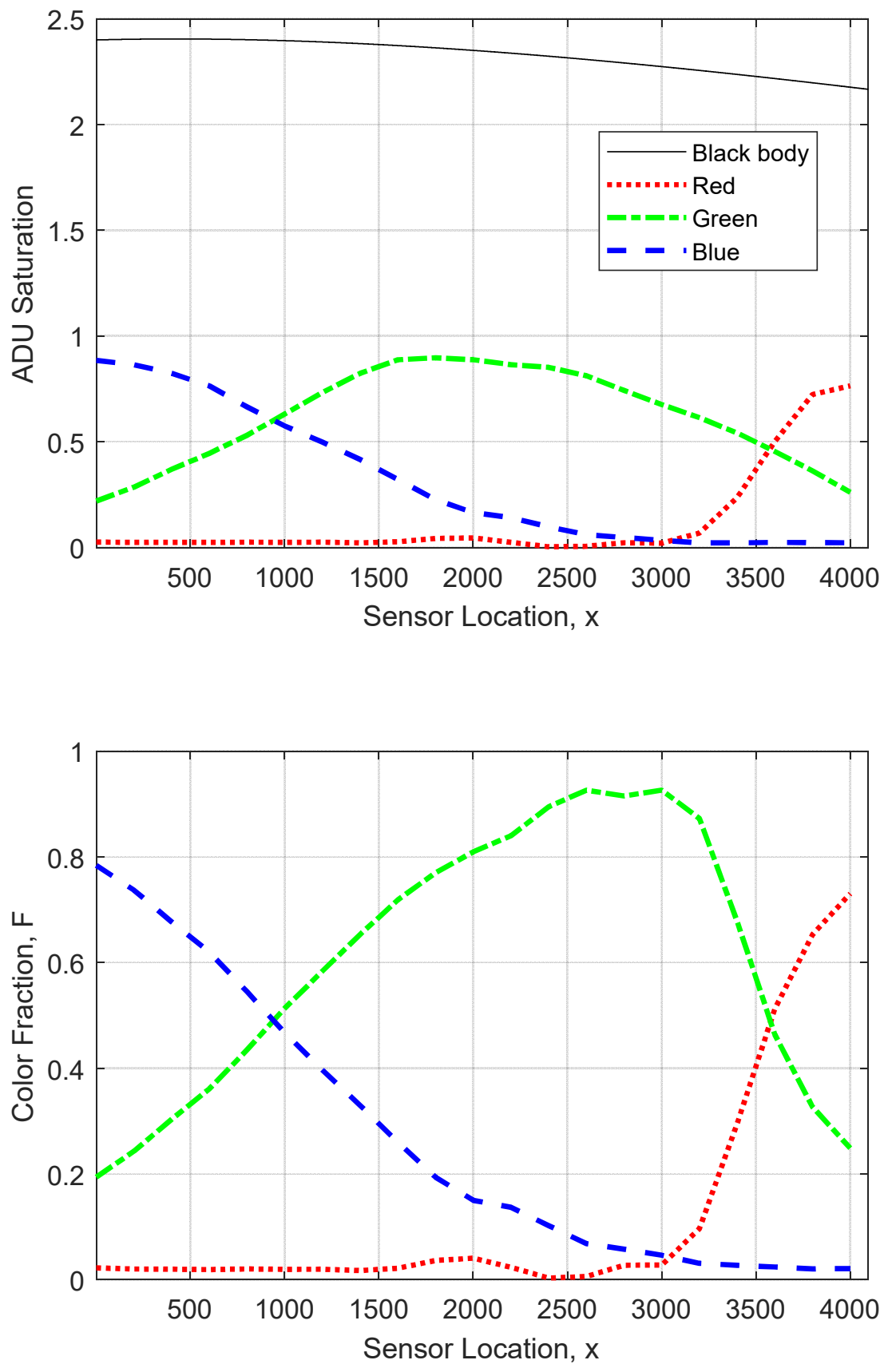

*FIG. 4. The translation of an observation into color fraction values. (Top panel) A simulated observation of a black body source and the measured intensities in red, green, and blue. The signal shown is normalized by the sensor's e⁻ well depth. The simulated exposure is adjusted so that the sensor is approximately 90% saturated, at most. (Bottom panel) The recorded signals of each color are converted to their fraction of the total signal for each position along the dispersion axis of the spectrometer. This is identical to Figure 3 except for the inclusion of noise and its measurement along sensor location instead of wavelength. The observation can now be related to sensor characteristics to create the wavelength solution.*

The matching of observed $F_C(x)$ to the characterized $F_C(\lambda)$ is repeated for each color at location x to arrive at N semi-independent value for the wavelength. Errors in the measured color fraction will result in errors in the deduced wavelength according to Equation 11, where $\frac{\delta\lambda}{\delta F}$ is the change in wavelength corresponding to a change in color fraction within the wavelength region of interest and $\sigma_F$ is from Equation 10.

$$\sigma_\lambda = \frac{\delta\lambda}{\delta F}\sigma_F \tag{11}$$

This is easily visualized in Figure 5, which shows $F_C(\lambda)$ for the green channel of a typical sensor. Steeper regions of the function lead to lower wavelength uncertainty for a given uncertainty in the color fraction measurement. An overall wavelength solution is eventually found for the spectrometer by referencing the known $F_C(\lambda)$ of each observed color at each location along the dispersion axis.

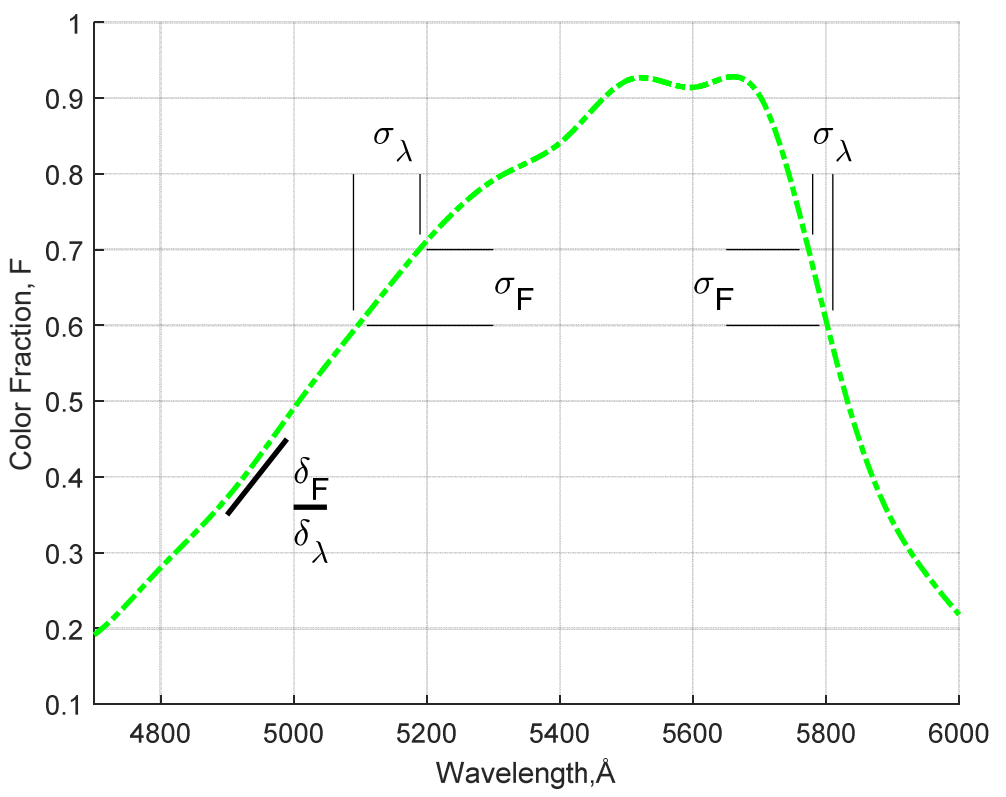

*FIG 5. The color fraction, F(λ), of a single color as a function of wavelength. An observed value of F(x) can be uniquely matched to a specific wavelength when the range is small enough to ensure a monotonic relationship. Uncertainty in the wavelength will depend on the uncertainty of F(x) and the slope of F(λ) in the region. Wavelength uncertainties are smallest in steep regions of color fraction.*

The user who wishes to predict or optimize performance must look to Equations 10 and 11 over their instrument's wavelength coverage. Specifically, the overall quality of the calibration will be a function of signal strength, the number of pixels along the dispersion axis, the number of color filters, and the QE functions of each color. The effect of signal strength is straightforward, with the ability to capture a strong flux via a large $e^-$ well depth leading to lower shot noise and improved precision. Similarly, a greater number of pixels to capture the spectrum statistically improves the development of the solution. Increasing the number of colors also generally increases the amount of data provided by an exposure. The interrelation between the various QE(λ)s is more complicated. Areas of high QE(λ) will have improved SNR for a given channel. But there must also be strong overlap in color sensitivities since this method depends on the ratio of at least two colors. From Equation 11 it is also clear that the precision of an individual pixel is best when the color fraction changes quickly with respect to changes in wavelength. Areas of low or plateaued $F_C(\lambda)$ will provide little useful information for a color. A combination of filters which provide overlapping regions of reasonable $F_C(\lambda)$ and steep changes in sensitivities to wavelength are expected to provide the best results. The details of optimization are left to future work. But these traits are seen in the test case used in the following simulation.

If the instrument is used to determine RV from Doppler shifts in wavelength, then a stellar spectrum is observed in association with the calibration. The location of features in the observed spectrum will appear shifted on the sensor in comparison to a reference spectrum due to either Doppler-induced changes in wavelength or a change in the spectrograph's wavelength solution. Calibration is intended to remove the effects of the latter.

The relative difference in RV between two observations is deduced by the observed shift in a feature's wavelength using Equation 12, where λ' is the new observation, $\lambda_o$ is the reference observation, and $\lambda_{rest}$ is the rest wavelength of the feature.

$$RV = \frac{(\lambda' - \lambda_o)c}{\lambda_{rest}} \tag{12}$$

Errors in the wavelength solution lead directly to errors in the determination of λ', and hence RV. A normal error in the mapping of sensor position to wavelength, $\sigma_\lambda$, quite directly leads to RV errors given by Equation 13. Discussing error in terms of RV instead of wavelength is more intuitive for this application and allows direct comparison of results obtained across the sensor since RV is expected to be a single value regardless of wavelength.

$$\sigma_{RV} = \frac{\sigma_\lambda c}{\lambda_{rest}} \tag{13}$$

In practice, the entire spectrum is used in RV determination. All of the features can be analyzed as in the case above, or more commonly, through the shift determination of large regions of the stellar spectrum relative to the reference using methods such as least squares alignment or cross correlation. The overall RV precision limit due to wavelength solution uncertainties using all pixels, p, and weighing each by its calculated precision is given below, where P is the total number of pixels.

$$\bar{\sigma}_{RV} = \sqrt{\frac{1}{\sum_{p=1}^{P} \sigma_{RV}^{-2}}} \tag{14}$$

To summarize, the proposed methodology for a color calibration system requires an initial sensor characterization using traditional techniques followed by a greatly simplified operational use. A sufficiently high-resolution spectrograph is coupled to the color-sensitive sensor and a spectral source such as ThAr is imaged to simultaneously capture features and colors. The wavelength solution of the sensor is determined using the location of known features and the relative intensities of colors are used to create $F_C(\lambda)$ for each type of filtered pixel. The same color-sensitive sensor is then paired to the spectrograph used for scientific purposes. Calibration exposures are taken using the instrument's primary optical path and may use any light source which emits brightly across the desired wavelength range. The observed color fraction for each pixel is calculated by comparing its ADU to other colors at the same dispersion location. The precision of each color fraction is calculated analytically. The observation is then numerically matched to a unique wavelength using the initial characterization's $F(\lambda)$. The overall wavelength solution is finally determined by combining all of the pixel-based results, weighed by the precision estimates. Scientific observations are finally made of the target through the same optical path and at nearly the same time.

## 3. Simulation

A simulation is now presented to demonstrate the use of this calibration method and to validate the error analysis of Section 2. The flux from a spectrally featureless light source is mathematically generated and passed through a modelled spectrograph. Each pixel's ADU value is converted to a color fraction using other colors at the same dispersion location. Each color fraction is then matched to the known response of the sensor to recover its wavelength. The quality of the recovered wavelength solution using color information is shown to agree with the analytical expectations.

The characteristics of the Kodak KAI-11002 color CCD are used in the simulation. This camera represents a high-end amateur product and allows abundant room for improvement by the professional. The camera is uncooled, has a well depth of ~60,000 $e^-$, a read noise of 30 $e^-$, a charge transfer efficiency greater than 0.99999, and a 14-bit ADC. The simulated calibration light source is the featureless spectrum of a 6000K blackbody with its intensity defined by Plank's Law. The simulation fixes the gain as the ratio of $2^{\text{bit depth}}$ to the $e^-$ well depth so that the maximum possible ADU is coincident with the full $e^-$ well depth. The flux of the calibration image is scaled such that the highest ADU reported by any pixel is approximately 90% of the full well depth.

To conduct the simulation, three distinct wavelength solutions are used: a reference, a distorted, and a recovered. A reference solution is defined that approximates the coverage of the spectrograph. A distorted solution is generated using the reference and is used to create the observation. This distorted solution is created randomly for each run of the simulation by permuting the reference solution. The recovered solution is the result found by applying color fraction techniques without knowledge of the distorted solution.

The reference solution is a fixed linear function having a coverage of 4700 Å at one edge and 7000 Å at the other. To create a uniquely distorted solution for each run, the corresponding wavelength responses for each edge of the sensor are perturbed independently by a random value up to a magnitude of 1 Å. A third pixel is randomly selected within the middle 2,000 pixels of the sensor as a crossover point having the same wavelength response as the linear reference. A second order polynomial is fit to the three locations to create the distorted wavelength solution used to simulate the observation. This imitates the type of challenging dispersion changes seen in spectrometers over time due to mechanical and optical changes (e.g. Thomas et al. 2016). The specific modelling of a distorted wavelength solution does not mimic the specific alterations seen in any particular instrument but provides a varied and unpredictable solution to be recovered. The deviation of the distorted solution from the linear reference is shown for five typical simulation passes in Figure 6. The observation is simulated by multiplying the calibration source's flux at each pixel's location by the QE(λ) for each type of filter. This is recorded in ADU, to include shot noise, read noise, and ADC rounding.

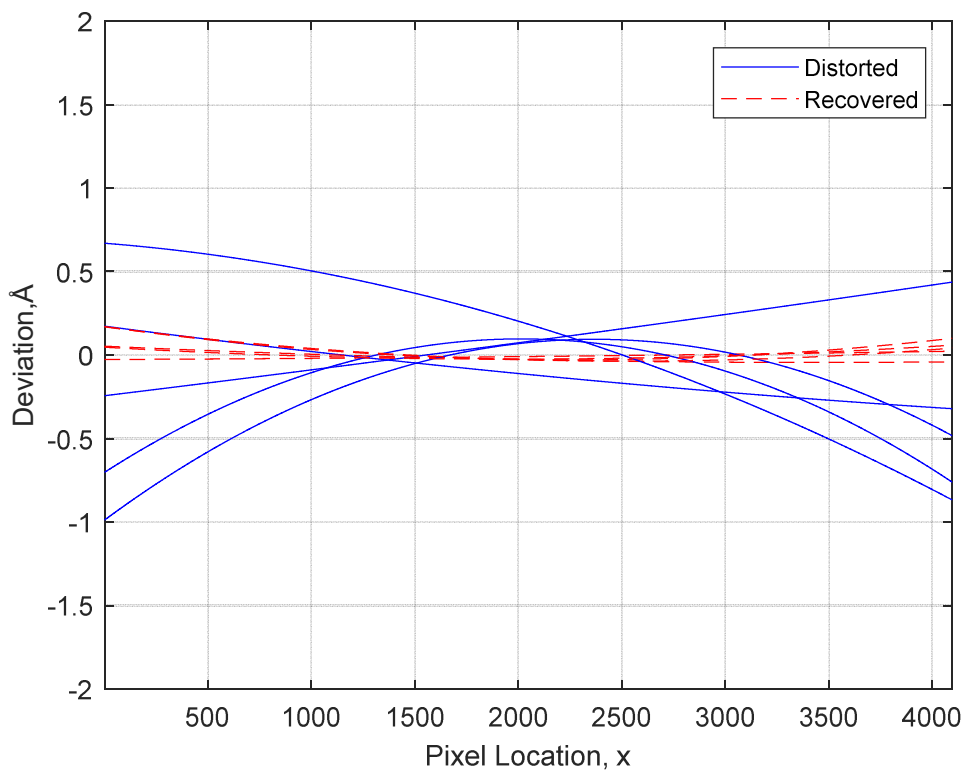

*FIG. 6. Typical deviations of the distorted wavelength solution used to create a simulated observation from the linear dispersion reference. Five typical examples are shown (solid). Error in the recovered wavelength solution for the same five simulations are also shown (dashed).*

The ADU reported by each pixel is converted to a color fraction using Equation 4. These results are then numerically matched to the sensor's known color response to calculate λ(x) for each pixel and its expected uncertainty using Equation 10. The linear reference is used as a starting guess in the iterative process of matching F(x) to the sensor's known F(λ). A unique wavelength solution for each pixel will be found so long as it is within the search range and F(λ) is monotonically increasing or decreasing within this range. A wavelength solution is generated from these results and compared to the distorted solution for an error analysis. The resolution of the spectrometer is not addressed because it does not directly impact this approach to calibration. Of course, resolution will be important to the scientific stellar observations, in which the shift of spectral features must be carefully measured (Bouchy et al. 2001).

The difference between the recovered and distorted solutions for each pixel in a single simulation is shown for each color in Figure 7, along with the 1σ boundary calculated using Equation 10 and sensor characteristics based on the KAI-11002.[3] The simulated results agree nicely with the predicted error boundaries. Uncertainty varies greatly across the CCD. But high-quality regions are found across the entire CCD when all colors are considered, as seen in the bottom panel of Figure 7, in which only the best color's result is shown at each pixel location. The predicted RV error resulting from estimated wavelength errors using Equation 14 is 1162 $ms^{-1}$.

The wavelength errors are converted into velocities for each pixel using Equation 12 and their weighted mean defines the overall perceived RV error introduced by calibration uncertainty. After 1000 runs of the simulation, each having a randomized distorted solution and noise, the standard deviation of RV errors is 1035 $ms^{-1}$. Results are in high agreement with analytical predictions. The errors of the wavelength solution corresponding to the same five simulations are also shown in Figure 6, demonstrating consistent convergence towards the distorted solution used to generate the observation.

[3] http://www.datasheet4u.com/datasheet-pdf/ONSemiconductor/KAI-11002/pdf.php?id=1077786

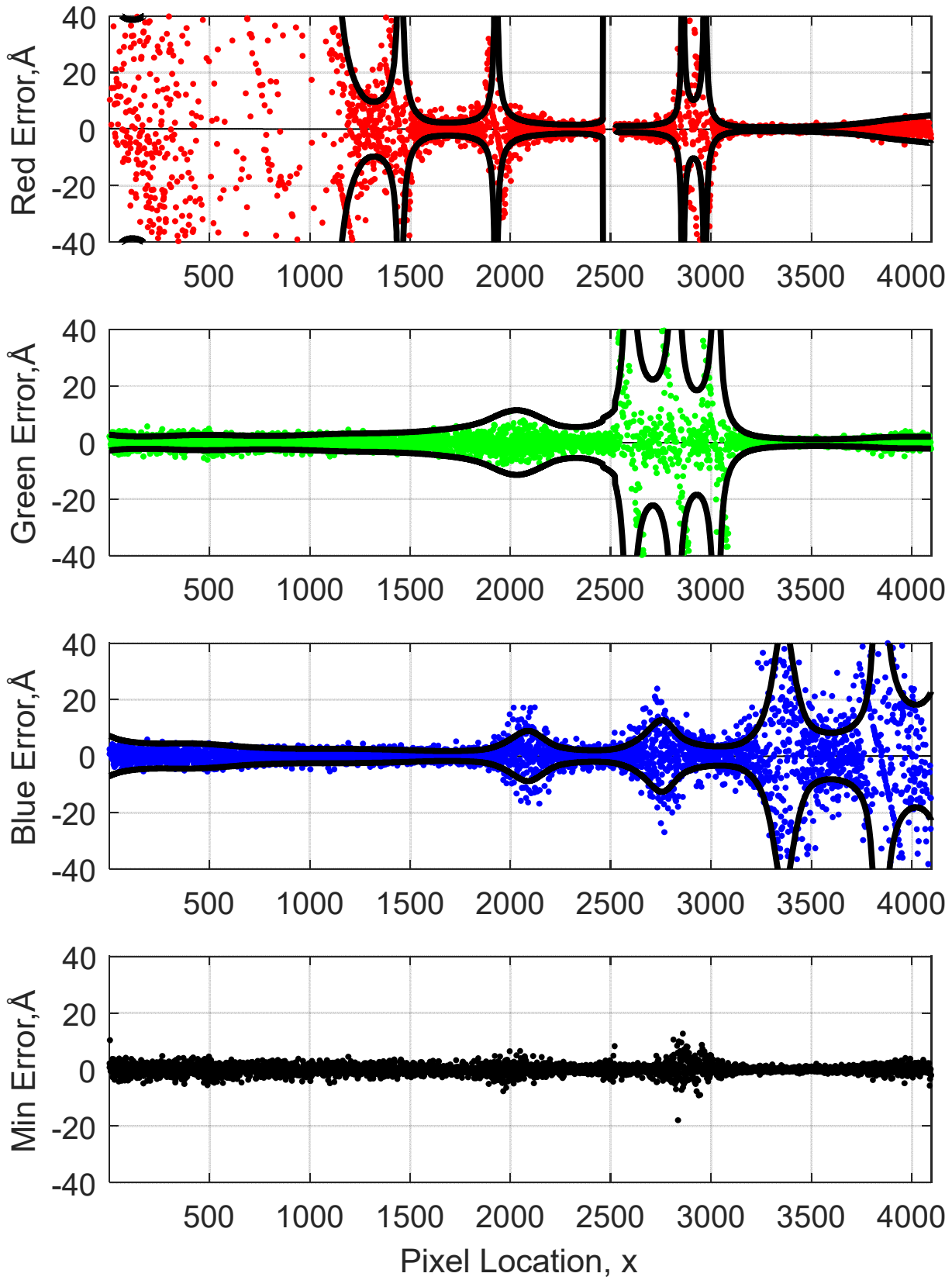

*FIG. 7. The error in wavelength for each color pixel in a simulated observation. The solid borders represent the predicted 1-σ error. Performance is very poor in regions where QE is very flat or QE is very low. Reasonable quality is maintained over the entire spectrum, however, when all three colors are considered. The bottom panel shows only the results for the lowest error color at each location. The regions of patterened error are the result of ADC rounding and low QE.*

As predicted earlier, the precision of wavelength calculations varies wildly for each color as a function of wavelength. This is because color fraction is defined as the ratio of the primary color's QE to the summation of all QEs at each dispersion location (Equations 2 and 10). Primarily, a particular color will yield very poor results in regions where it has low QE because the SNR will decrease accordingly. Uncertainty will also become large when only one color has usably high QE. The color fraction will approach unity in these cases and provide little useful information in wavelength determination. Although a more detailed analysis of optimized filter construction is beyond the scope of this paper, the simulated example here demonstrates that it is readily possible to construct a set of filters which provide usable results from at least one of the colors over a desired wavelength range.

The precision in this example is comparable to that of some large-scale surveys such as LAMOST (Wang & Lou 2012) but is significantly inferior to that obtained by the most precise modern instruments, which is on the order of 1 $ms^{-1}$ (e.g. Queloz et al. 2001). These results, however, are based on amateur OTS equipment. Consider a few straightforward approaches to improvement. The uncertainty of color fraction data for a specific sensor is inversely proportional to the SNR of the signal, given the assumptions used to create Equation 10 and realizing that SNR is the square root of the signal when only shot noise is considered. Since a calibration lamp can be arbitrarily bright, the SNR is constrained in practice by the maximum number of electrons that a pixel can collect before saturation. SNR at each location can be improved by increasing the e- well depth or increasing the number of pixels used to observe each dispersion location. Higher performance CCDs can provide $e^-$ well depths of 350,000[4], improving SNR in this example by a factor of 5.8. Secondly, taking four calibration exposures doubles the SNR again. Lastly, we have assumed that the spectrograph has only one pixel of each color at each location in the dispersion direction. A spectrum can easily be 60 pixels wide in the cross-dispersion axis and allow for 20 pixels of each color at each wavelength sampling. These three improvements increase the SNR by 464 and reduce the error by a factor of 21.5, to well under 100 $ms^{-1}$. Such a level of precision is fully capable of permitting the detection of brown dwarfs and giant planets, all without the complications associated with advanced calibration sources and separate optical paths. These improvements are speculative, of course, because they assume a system limited by white noise and free of systematics.

The simulation is repeated for increasing SNRs with the results shown in Figure 8. This is accomplished in the simulation by increasing $e^-$ well depth. But SNR can effectively be increased in a combination of ways discussed above. The predicted performance of various available sensors is marked. Their performance is also predicted when a calibration consists of four combined exposures and a spectrum wide enough to permit 20 pixels of each color at each x-location. Precision can be improved arbitrarily in this analysis. Of course, the effects neglected by Equation 10 such as read noise, variations among pixels (QE, size, and location), and the quality of the QE(λ) solution will eventually become limiting factors.

[4] http://www.andor.com/scientific-cameras/ikon-xl-and-ikon-large-ccd-series/ikon-xl-231

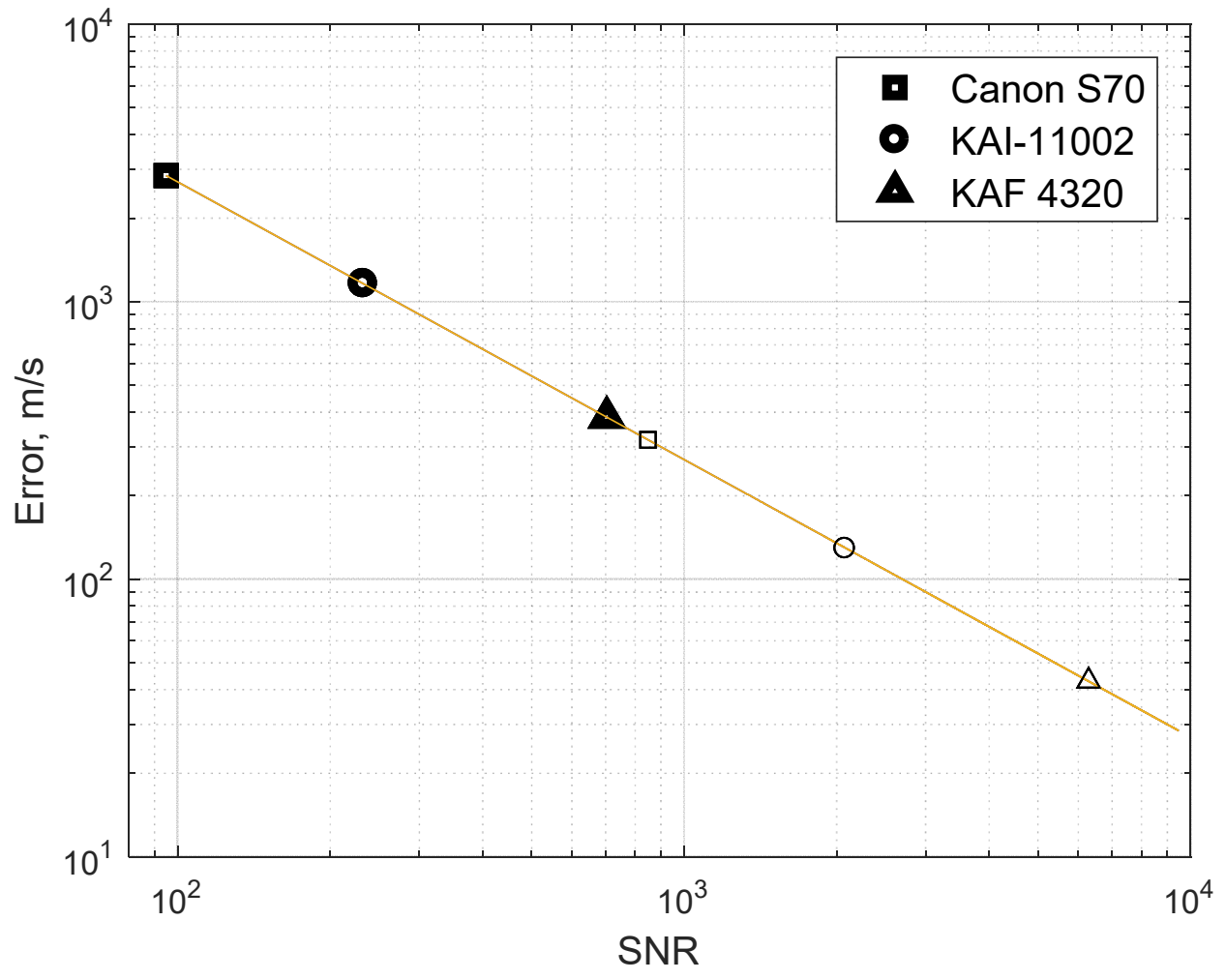

*FIG. 8. RV errors introduced by calibration error at various SNRs. Bold symbols represent the performance of existing sensors with a single exposure taken at 90% well depth saturation and using only one pixel of each color at each dispersion location. The Canon S70 is an obsolete point and shoot camera. The KAI-11002 is for high-end amateur astronomy use. The KAF 4320 is professional grade. The fainter symbols show performance for the same cameras when four exposures are combined, and a realistic spectrum width is incorporated.*

## 4. Discussion

Existing calibration methods are effective but require the use of dedicated optics and a specialized calibration source, as well as careful processing to avoid systematic errors. Although this new method of color calibration is only introduced through theory and simulation, it may offer a simpler approach.

The primary limit to the precision of color calibration is the achievable signal of each pixel. This can be increased by selecting a sensor with a higher $e^-$ well depth, combining multiple exposures, or increasing the cross-dispersion width of the projected spectrum. Noise is least detrimental in spectral regions where the change in $F(\lambda)$ is the steepest. This study examined an existing sensor with red, green, and blue filters. Performance may be improved by selecting narrower bandpass filters and using more than three colors, although developing customized sensors would add a new level of expense.

The reader might imagine applications outside of the visible spectrum where exotic calibration sources are available in the lab but impractical for long-term use in the field. The need for calibration spectra only during initial sensor characterization would be highly advantageous in such cases. Only a non-specific light source which simply emits flux in the wavelength range of interest would be required for operational calibration.

This method also offers the potential for self-calibrating stellar exposures with no need for distinct calibration. Spectral features move due to Doppler shifts but the dispersion of the spectrograph does not. A feature may become bluer, for instance, and shift to the left. But the location of a particular shade of blue on the sensor is unaffected by the target's nature. The movement of a feature can be used to detect its Doppler shift while the colors of the same spectrum are used to develop the wavelength solution. This would be a very pure form of calibration because it allows no variation in time, optics, or spectral characteristics between the scientific observation and the calibration. Noise will likely be increased, however, as compared to the use of a bright calibration lamp.

Not fully addressed in this paper, is the criticality of precisely characterizing the sensor's $F(\lambda)$ for each color. The researcher will likely need to fully develop $F(\lambda)$ for each color filter using a traditional source such as Iodine or ThAr. Once this is done, however, it is defined for all the pixels of each color and not affected in the future by spectrograph changes. This characterization would only need to be done once for a given sensor, assuming that the qualities of the filters do not change with time or environmental factors. While the scientific results obtained from stellar observations will depend on the resolution and characteristics of the spectrograph used at the telescope, the color calibration itself does not depend explicitly on the resolution of the spectrograph.

This approach could also be applied by sequentially inserting color filters into the optical path. This would have the advantage of being more readily accomplished with existing monochromatic systems. Also, using distinct color-filtered calibration exposures allows for the scientific observation to be unfiltered and shorter. Lastly, there would be less concern that flux variations in the continuum between adjacent pixels of different colors could introduce error since each pixel will provide all of its own color information. But this approach negates a bit of the elegance in using a color sensor to eliminate all machinery used for calibration and the collection of all calibration data in a single exposure. Experimentation with sequential filtering is recommended for the professional team that already has an operating monochromatic spectrometer. The color sensor approach is recommended for the budget-minded experimenter or for applications where weight and complexity must be kept to a minimum, such as in satellites or unmanned aircraft.

## 5. Conclusion

The analysis and simulation of a new method of spectral calibration using color sensors has shown the potential to provide calibration with reduced instrumentation and simplified processing. Calibration is performed with no specialized light source and no optics other than those used for the scientific observation. The method is limited by the quality of the initial characterization of the sensor's color filters. Recommendations for experimentation using either color sensors or color filters have been presented. Positive results from experimentation may lead to a more widely available means of calibrated spectroscopy for radial velocity studies or in other fields where simplicity and compactness is critical.

**ACKNOWLEDGMENTS**

This work was supported through numerous accommodations by the 738th Air Expeditionary Advisory Group, Kandahar, Afghanistan, where the work was performed. Special thanks are extended to Manny Fiterre, Trent Alexander, Hanif Rahimi, and Carolynne Thomas for many spirited conversations concerning this topic. Many thanks are due to Dr. Jian Ge and his exoplanet team at the University of Florida as well as the faculty of the US Air Force Academy's Astronautical Engineering department for their technical advice. Finally, the very thorough inputs from a very expert anonymous reviewer during this paper's refinement were absolutely critical.

## REFERENCES

Bouchy, F., Pepe, F., & Queloz, D. 2001, A&A, 374, 733

Butler, R., Marcy, G., Williams, E., et al. 1996, PASP, v.108, p.500

Ge, J., Lee, B., de Lee, N., et al. 2009, SPIE, Vol. 74400L, 10pp.

Grieves, N., Ge, J., Thomas, N., et al. 2017, submitted to ApJ.

Lovis, C. & Pepe, F. 2007, A&A, 468, 1115

Marcy, G. & Butler, P. 1992, PASP 104, 270-277

Mayor, M., Pepe, F., Queloz, D., et al. 2003, The Messenger 114, 20

Pepe, F., Cristiani, S., Rebolo, L., et al. 2010, SPIE, Vol. 7735, id. 77350F

Queloz D., Mayor, M., Udry, S., et al. 2001, The Messenger 105:1-7

Thomas, N., Ge, J., Grieves, N., et al. 2016, PASP, 128,192

Wang, F. & Luo, A. 2012, ASI Conference Series, Vol. 6, pp 253-256

Wildi, F., Pepe, F., Chazelas, G., et al. 2010, Proc. SPIE 7735, Ground-based and Airborne Instrumentation for Astronomy III, 77354X

Wright, J. T. 2017, Radial Velocities as an Exoplanet Discovery Method, arXive: 1707.07983v2